\documentclass[a4paper]{aa}

\usepackage{graphicx,float}
\usepackage{latexsym,amsmath,amssymb}
\usepackage{natbib}
\bibpunct{(}{)}{;}{a}{}{,}

\def \arvind 	{A.~N. Parmar}
\def \didier 	{D. Barret}
\def \josh 	{J.~E. Grindlay}
\def \laurence 	{L. Boirin}
\def \stephane  {S. Paltani}

\def \estec {Astrophysics Missions Division, Research and Scientific Support
		Department of ESA, ESTEC, Postbus 299, NL-2200 AG
		Noordwijk, The Netherlands}

\def \cfa	{Harvard-Smithsonian Center for Astrophysics, 60 Garden Street
		Cambridge, MA 02138, USA}

\def \cesr	{Centre d'Etude Spatiale des Rayonnements, CNRS/UPS, 
		9 Av. du Colonel Roche, 31028 Toulouse Cedex 4, France}

\def \marseille 	{Laboratoire d'Astrophysique de Marseille, Traverse du
		Siphon, BP 8, 13376 Marseille Cedex 12, France }

\newcounter{mycountera}
\newcommand {\numaup} {$^{\themycountera}$ \addtocounter{mycountera}{1}}

\newcounter{mycounterb}
\newcommand {\numbup} {$^{\themycounterb}$ \addtocounter{mycounterb}{1}}

\def \ang {$\rm\AA$}
\def\ergsec{\hbox{erg s$^{-1}$}}

\def\countsec{\hbox{counts s$^{-1}$}}

\def \rsun {\ifmmode$R$_{\odot}\else R$_{\odot}$\fi}
\def \nh {$N{\rm _H}$}
\def \nhdbb {$N{\rm _H^{DBB}}$}
\def \nhpl {$N{\rm _H^{PL}}$}
\def \nhgal {$N{\rm _H^{Gal}}$}

\def \xiunit {\hbox{erg cm s$^{-1}$}}

\def \hcm {\hbox {\ifmmode $ atoms cm$^{-2}\else atoms cm$^{-2}$\fi}}

\def\approxgt{\mathrel{\hbox{\rlap{\lower.55ex \hbox {$\sim$}}
        \kern-.3em \raise.4ex \hbox{$>$}}}}
\def\approxlt{\mathrel{\hbox{\rlap{\lower.55ex \hbox {$\sim$}}
        \kern-.3em \raise.4ex \hbox{$<$}}}}

\newcommand {\degree} {$^{\circ}$}

\def\arcsec{\hbox{$^{\prime\prime}$}}

\newcommand {\chisq} {$\chi ^{2}$}
\newcommand {\rchisq} {$\chi_{\nu} ^{2}$}

\newcommand {\phind} {$\alpha$} 
\newcommand {\ecut} {$E_{\rm c}$}

\newcommand {\ttnh} {$\times~$10$^{22}$~atom~cm$^{-2}$}

\newcommand {\eline} {$E_{\rm line}$}
\newcommand {\eedge} {$E_{\rm edge}$}

\newcommand {\ktin} {$kT_{\rm in}$}
\newcommand {\kevnorm} {ph.~keV$^{-1}$~cm$^{-2}$~s$^{-1}$}
\newcommand {\sig} {$\sigma$}
\newcommand {\ew} {$EW$}
\newcommand {\ews} {$EW$s}

\newcommand {\fetfour} {\ion{Fe}{xxiv}}
\newcommand {\fetfive} {\ion{Fe}{xxv}}
\newcommand {\fetsix} {\ion{Fe}{xxvi}}
\newcommand {\nitseven} {\ion{Ni}{xxvii}}
\newcommand {\ssixteen} {\ion{S}{xvi}}

\newcommand {\ka} {K$\alpha$}
\newcommand {\kb} {K$\beta$}

\def \nineteen {XB\,1916-053}
\def \mxb {MXB\,1658-298}
\def \bigdip {X\,1624-490}
\def \twelve {X\,1254-690}
\def \grs {GRS\,1915+105}
\def \gro {GRO\,J1655-40}
\def \gx {GX\,13+1}
\def \cir {Cir\,X-1}
\def \exo {EXO\,0748-676}
\def \eighteentt {4U\,1822-371}

\begin{document}

\title{Discovery of X-ray absorption features from the dipping low-mass X-ray binary \nineteen\ with XMM-Newton}

\author{\laurence\inst{1} \and \arvind\inst{1} \and \didier\inst{2} \and \stephane\inst{3} \and \josh\inst{4}}

\offprints{L. Boirin, \email{L.Boirin@sron.nl}}

\institute{\estec \and \cesr \and \marseille \and \cfa}

\date{Received 21 Oct 2003 / Accepted 03 Feb 2004}

\authorrunning{L. Boirin et al.}

\titlerunning{XMM-Newton observation of \nineteen}

\abstract{We report the discovery of narrow  \fetfive\
and \fetsix\ \ka\ X-ray absorption lines at 6.65$\,^{+0.05}_{-0.02}$
and 6.95$\,^{+0.05}_{-0.04}$~keV in the persistent emission of the
dipping low-mass X-ray binary (LMXB) \nineteen\ during an XMM-Newton
observation performed in September 2002.  In addition,
there is marginal evidence for absorption features at 1.48~keV,
2.67~kev, 7.82~keV and 8.29~keV consistent with \ion{Mg}{xii},
\ssixteen, \nitseven\ \ka\ and
\fetsix\ \kb\ transitions, respectively.  Such absorption lines from
highly ionized ions are now observed in a number of high inclination
(ie. close to edge-on) LMXBs, such as \nineteen, where the inclination
is estimated to be between 60--80\degree.  This, together with the
lack of any orbital phase dependence of the features (except during
dips), suggests that the highly ionized plasma responsible for the
absorption lines is located in a cylindrical geometry around the
compact object. Using the ratio of \fetfive\ and
\fetsix\ column densities, we estimate the photo-ionization 
parameter of the absorbing material, $\xi$, to be $10^{3.92}$
\xiunit. Only the \fetfive\ line is observed during dipping intervals
and the upper-limits to the \fetsix\ column density are consistent
with a decrease in the amount of ionization during dipping intervals.
This implies the presence of cooler material in the line of sight
during dipping.  We also report the discovery of a 0.98~keV absorption
edge in the persistent emission spectrum.  The edge energy decreases
to 0.87~keV during deep dipping intervals. The detected feature may
result from edges of moderately ionized Ne and/or Fe with the average
ionization level decreasing from persistent emission to deep
dipping. This is again consistent with the presence of cooler material
in the line of sight during dipping.  }

\maketitle

\section{Introduction}
\label{sec:intro}

\nineteen\ is a LMXB showing periodic intensity dips in its light
curve \citep{1916:walter82apjl,1916:white82apjl}. Dips are believed to
be due to obscuration of the central X-ray source by a vertical
structure located at the outer edge of the accretion disk and
resulting from the impact of the accretion flow from the companion
star into the disk \citep{1916:white82apjl}.  The presence of dips in
\nineteen\ and the lack of X-ray eclipses from the companion star
indicate that the system is viewed relatively close to edge-on, at an
inclination angle in the range $\sim$60--80\degree\
\citep{frank87aa,1916:smale88mnras}.  The X-ray dip period is 50
minutes \citep[e.g., ][]{1916:white82apjl}, the
shortest amongst the dipping sources.  The optical counterpart of
\nineteen\ shows a modulation with a period $\sim$1~\% longer than the
X-ray dip period
\citep[e.g., ][]{1916:callanan95pasj}. 
This discrepancy has led to several interpretations including superhumps 
\citep{1916:schmidtke88aj} and a hierarchical triple system model \citep{1916:grindlay88apjl}. \citet{1916:retter02mnras} recently favored the superhump model, which
invokes a precessing accretion disk, and which identifies the X-ray
period as orbital.
\nineteen\ is a  type~I X-ray burster \citep{1916:becker77apjl}, 
indicating that the compact object is a neutron star.
\nineteen\ has shown quasi-periodic oscillations  at 
various frequencies ranging from $\sim $0.2 to 1300 Hz
\citep{1916:boirin00aa}, and a 270~Hz highly coherent
oscillation during an X-ray burst
\citep{1916:galloway01apjl}.

\nineteen, as most dippers, shows spectral hardening 
during dipping. However, the spectral evolution is not consistent with
a simple increase of photo-electric absorption by cool material, as an
excess of photons is present at low energy. Two approaches have been
proposed for modelling the spectral evolution during dips. In the
``absorbed plus unabsorbed'' approach \citep[e.g.,
][]{0748:parmar86apj}, the persistent (non-dipping) model is used to
fit the intensity-selected dip spectra, but is divided into two
parts. One part is allowed to be absorbed, whereas the other one is
not. The spectral evolution during dipping is well accounted for by a
large increase in the column density of the absorbed component, and a
decrease of the normalization of the unabsorbed component. The latter
decrease has been attributed to effects of electron scattering in the
absorber.  In the ``progressive covering'', or ``two-component'' or
``complex continuum'' approach
\citep[e.g., ][]{1916:church97apj}, the X-ray emission originates from
two components. The first one, generally modelled as a blackbody, is
from a point-like region, such as a boundary layer around the compact
object.  The second component, generally modelled as a power-law, comes
from an extended region, such as an accretion disk corona. The complex
continuum approach allows both the persistent and the various
intensity-selected dipping spectra to be modelled with the same two
components, and explains the spectral changes by allowing partial and
progressive covering of the extended source by an opaque absorber that
occults various fractions of the source. The absorption applied to the
point-source component is allowed to vary independently of the
absorption applied to the extended component, but no partial covering
is needed as the source is supposed to be point-like and thus fully
covered by the absorber in the line of sight during the dips. Both
approaches have been applied to \nineteen\ \citep[e.g., ][
respectively]{1916:yoshida95pasj,1916:church97apj}.

In addition to the continuum emission, a broad emission feature
interpreted as fluorescent line emission from neutral Fe was detected
from \nineteen, at 5.60$^{+0.53}_{-0.43}$~keV \citep{1916:smale92apj},
6.14$^{+0.18}_{-1.07}$~keV \citep{1916:bloser00apj} and
5.9$^{+0.2}_{-0.1}$~keV \citep{asai00apjs}.  The low energy of the
emission lines compared to the 6.4~keV expected for neutral Fe led
\citet{1624:parmar02aa} to propose that the broad emission feature was
modulated with $\sim$7~keV absorption features, which were not
included in the spectral model, reducing the measured energy of the
emission feature. Narrow absorption features from highly ionized Fe
and other metals are now observed in a growing number of LMXBs (see
Table~\ref{tab:sources}), thanks to the improved sensitivity and
spectral resolution offered by the new generation of instruments
on-board of {\it Chandra} and XMM-Newton.  These discoveries indicate
that a highly ionized plasma is present in these systems, that was so
far not taken into account in models. Therefore, the study of
these lines appears to be extremely important to characterize the
geometry and physical properties of this plasma, that could be a
common property of accreting systems. The presence of absorption lines
could be related to the viewing angle of the system.

In this paper, we report the discovery of narrow \fetfive\ and
\fetsix\ \ka\ X-ray absorption lines near 7~keV in the dipping LMXB
\nineteen. Absorption features
due to highly ionized Mg, S and Ni may also be present. The \fetfive\
line is also observed during dipping intervals.  We also report the
discovery of an absorption edge at an energy of 0.98~keV and 0.87~keV
during the persistent and dipping intervals, respectively.  Detailed
modelling of the continuum emission, comparison of the ``absorbed plus
unabsorbed'' and ``complex continuum'' approaches, and discussion of
their physical interpretations are the subject of a separate paper
(Webb et al., in preparation), and are therefore not
presented here. In this paper, we focus on the X-ray absorption
features, and, in order to extract their properties, we fit the
continuum with a simple model and follow the complex continuum
approach.

\section{Observation and data reduction}

The XMM-Newton Observatory \citep{jansen01aa} includes three
1500~cm$^2$ X-ray telescopes each with an European Photon Imaging
Camera (EPIC, 0.1--15~keV) at the focus.  Two of the EPIC imaging
spectrometers use MOS CCDs \citep{turner01aa} and one uses pn CCDs
\citep{struder01aa}. Reflection Grating Spectrometers \citep[RGS,
0.35--2.5~keV,][]{denherder01aa} are located behind two of the
telescopes.  \nineteen\ was observed by XMM-Newton for 17~ks on
September 25, 2002, from 03:55 to 08:31 UTC.  All the EPIC cameras
were operating in timing mode, with medium thickness optical blocking
filters applied.  As the pn CCD is more sensitive to the presence of
lines than the MOS CCDs with an effective area a factor 5 higher at
$\sim$7~keV and a better energy resolution, we concentrate on the
analysis of pn data.  We also carry out a spectral analysis of RGS
data.

We used the X-ray data products generated by the Pipeline Processing
Subsystem on October 2002. We further filtered these products and
generated lightcurves and spectra using the Science Analysis Software
(SAS) version 5.4.1. Electronic noise and hot or flickering pixels
were rejected. No high background intervals were present in the data.

\begin{figure*}[!ht]
\centerline{\includegraphics[angle=90,width=\textwidth]{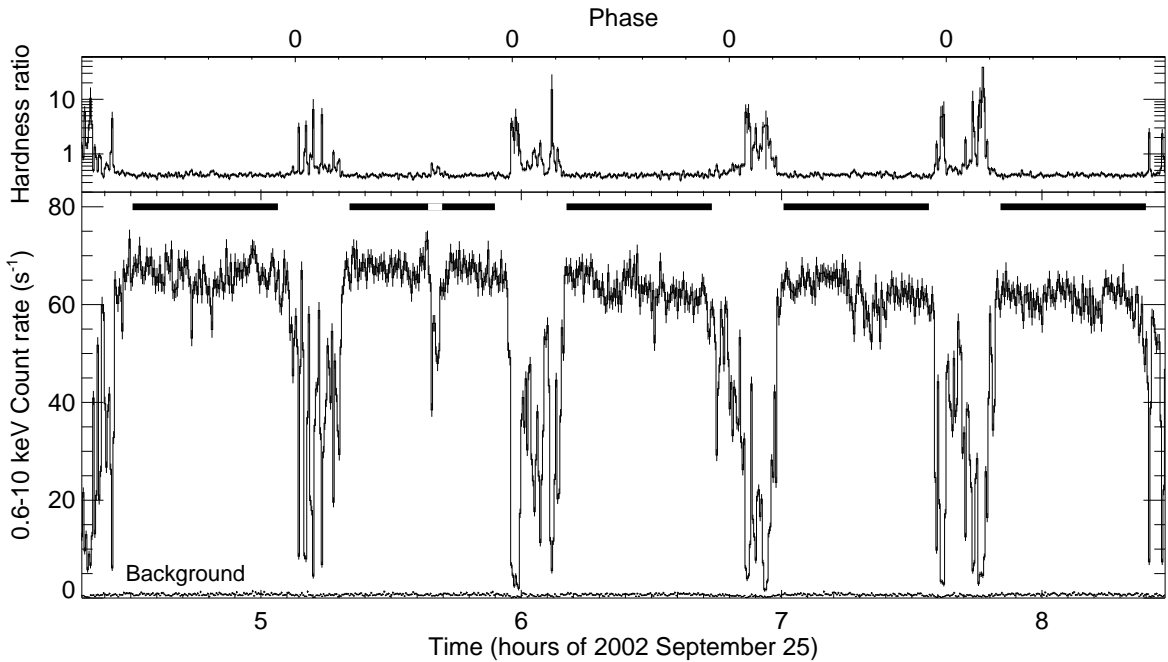}}
\caption{0.6--10~keV EPIC pn background-subtracted light curve of
\nineteen\ on September 25, 2002, with a binning of 20~s (lower panel)
showing parts of 6 deep dips and weaker intermediate dipping
activity. Times are barycenter-corrected.  The 0.6--10~keV background
light curve taken from a region adjacent to the source is also
shown. The upper panel shows the hardness ratio (counts in the
2.5--10~keV band divided by counts in the 0.6--2.5~keV band) with a
logarithmic scale, and a binning of 20~s. The phase according to the
ephemeris of \citet{1916:chou01apj} is indicated on the top axis. The
thick horizontal lines indicate the intervals used for the persistent
emission spectral analysis, corresponding to phases 0.25--0.92, and
excluding the secondary dipping activity near 5.7~h.}
\label{fig_pn_lc}
\end{figure*}

In pn timing mode, only one pn CCD chip (corresponding to a field of
view of 13\farcm6$\times$4\farcm4) is used and the data from that chip
are collapsed into a one-dimensional row (4\farcm4) to be read out at
high speed, the second dimension being replaced by timing information.
This allows a time resolution of 30~$\mu$s, and photon pile-up occurs
only for count rates above 1500~\countsec. We selected only single and
double events (patterns 0 to 4), and extracted source events from a
53\arcsec\ wide column centered on the source position. Background
events were obtained from a column of the same width, but centered
115\arcsec\ from \nineteen.  We used the latest response matrix file
for the pn timing mode provided by the XMM-Newton calibration team
(epn\_ti40\_sdY9.rsp, released in 2003 January).  We generated an
ancillary response file using the ftools {\tt arfgen}.  In the source
spectrum, we observe two strong emission features near 0.25 and
0.45~keV which are not removed by spectral modelling. Such deviations
have been reported in the pn timing mode spectrum of XTE\;J1751-305 by
\citet{1751:miller03apjl}, who suggest that these are due to incorrect
modelling of the instrumental response. To check this, we have
downloaded several data sets from the XMM-Newton archives and
extracted pn timing mode spectra for several X-ray binaries. For all
of them, we find features below 0.6~keV similar to those observed in
the spectrum of \nineteen. Thus, we conclude that these features have
an instrumental origin.  In the inspected spectra, we also note
deviations to the continuum fit above 10~keV.  We therefore examine pn
timing mode data in the 0.6--10~keV energy range.  The EPIC pn spectra
were rebinned to oversample the full width at half maximum of the
energy resolution by a factor 3, and to have a minimum of 20 counts
per bin to allow the use of the $\chi^2$ statistic.  In order to
account for systematic effects a 2\% uncertainty was added
quadratically to each spectral bin.

The SAS task {\tt rgsproc} was used to produce calibrated RGS event
lists, spectra, and response matrices.  The RGS spectra were examined
in the 0.5--2.0~keV range, where the source count rate is high
compared to the background count rate.  The spectra were rebinned to
have a minimum of 20 counts per bin to allow the use of the $\chi^2$
statistic.

Modelling of EPIC and RGS spectra was carried out using XSPEC version
11.2.  The photo-electric absorption cross sections of
\citet{morrison83apj} are used throughout.  All spectral uncertainties
are given at 90\% confidence, and upper limits at 95\% confidence.

\section{Results}

\subsection{X-ray lightcurve}

The 0.6-10~keV background-subtracted light curve obtained from the
EPIC pn is shown in Fig.~\ref{fig_pn_lc} (bottom panel) with a binning
of 20~s. The upper panel shows the hardness ratio (counts in the
2.5-10~keV band divided by counts in the 0.6-2.5~keV band).  Times
were barycenter-corrected using the SAS task {\tt barycen}.  The X-ray
phase is indicated on the top horizontal axis of Fig.~\ref{fig_pn_lc},
and computed using MJD~50123.00944 as zero-phase epoch and an X-ray
folding period of 3000.6508~s. This X-ray ephemeris was obtained by
\citet{1916:chou01apj} from a timing analysis of RXTE observations
carried out between  February and October 1996 combined with historical
X-ray data obtained from 1976 to 1996. Dip centers are expected at
phase 0.

\begin{figure}[!h]
\centerline{\includegraphics[angle=90,width=0.5\textwidth]{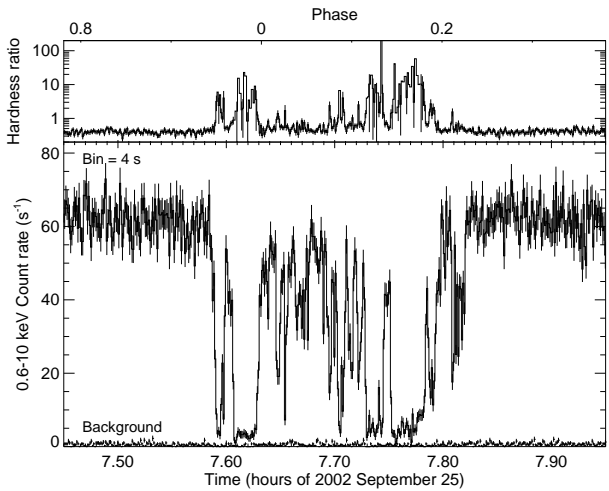}}
\caption[]{A 0.6--10~keV pn background-subtracted lightcurve of the
dip seen on September 25, 2002, at $\sim$7.7~hours plotted with a time
resolution of 4~s, showing the rapid intensity variability.  Times are
barycenter-corrected. The 0.6--10~keV background light curve taken
from region adjacent to the source is also shown. The upper panel
shows the hardness ratio (counts between 2.5--10~keV divided by those
between 0.6--2.5~keV). The phase according to the
\citet{1916:chou01apj} ephemeris is indicated.}
\label{fig:zoom_dip}
\end{figure}

No X-ray burst was observed during the $\sim$4~hour XMM-Newton
observation. Four complete dips are visible, as well as parts of dips
at the start and at the end of the observation. Dip centers occur at
phase~$\sim$0.1, rather than phase~0. Such a phase jitter is
consistent with that observed so far from \nineteen: the 101 dip
centers recorded by \citet{1916:chou01apj} follow a distribution with
a mean phase of 0.00098 and a 1$\sigma$ fluctuation of 0.06.  The
dipping activity is associated with spectral hardening. This behavior
is usual for \nineteen\ and for most of the dippers. The dip shape
varies from dip to dip. The dipping intensity shows erratic
variability, ranging between $\sim$7\% and $\sim$85\% of the
persistent level. Fig.~\ref{fig:zoom_dip} shows an expanded view of
the fourth complete dip observed with a time resolution of 4~s.  At
times, the dipping is almost total in the 0.6--10~keV energy range
when plotted with this time resolution.  Secondary dipping activity
occuring at phase~$\sim$0.6 (i.e. $\sim$0.5 away from the dip centers)
is also visible, especially at $\sim$5.65~h.  The 0.6--10~keV
persistent intensity (outside the dips) decreased slowly throughout
the observation from $\sim$67 to $\sim$62~\countsec.

\begin{figure*}[ht!]
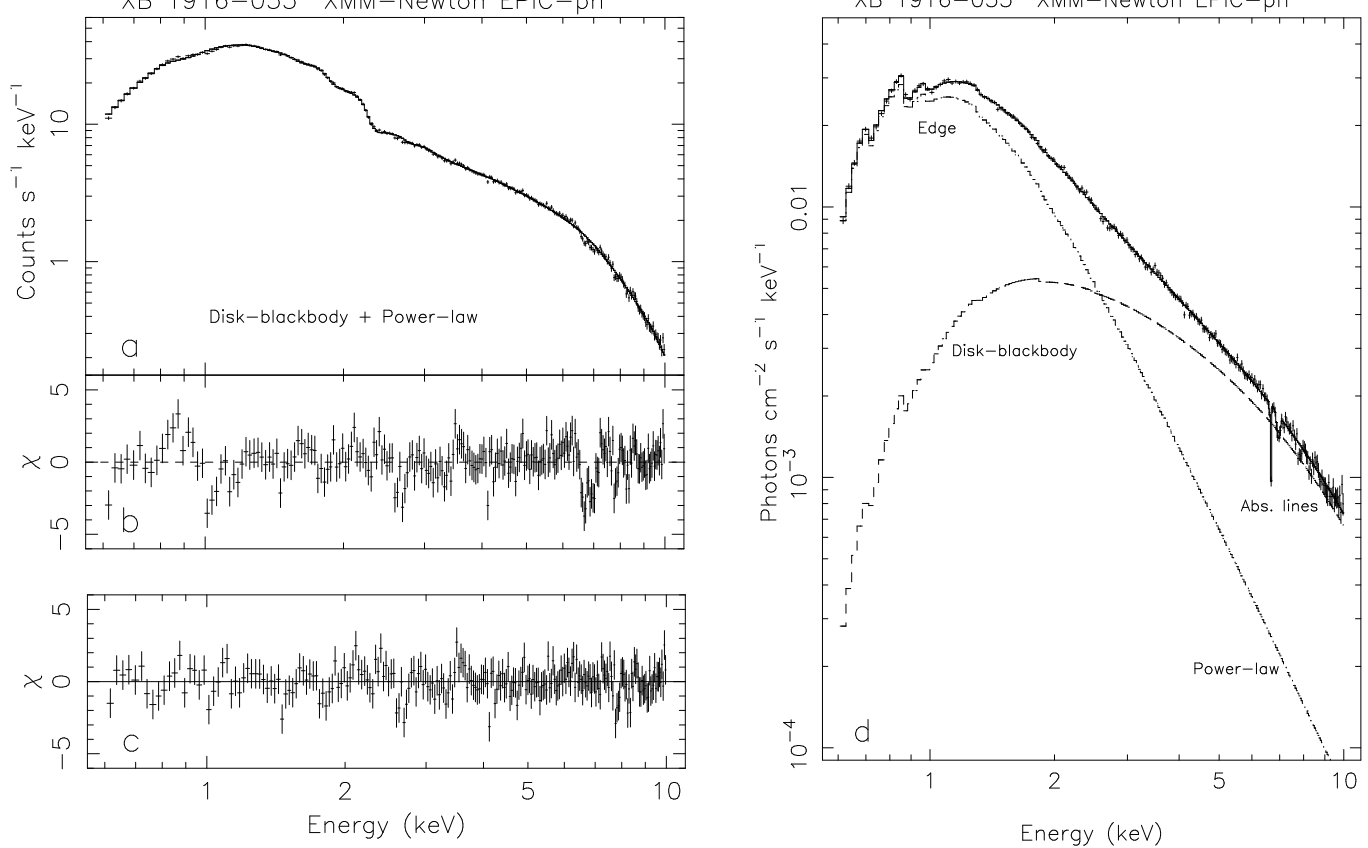

\includegraphics[angle=-90,width=0.5\textwidth]{0550f3ab.ps}
\hfill
\includegraphics[angle=-90,width=0.45\textwidth]{0550f3d.ps}\\

{\vspace{-3.8cm}\hspace{0.2cm}\includegraphics[angle=-90,width=0.49\textwidth]{0550f3c.ps}}\\
\caption{(a) pn spectra of the \nineteen\
persistent emission and the best-fit disk-blackbody plus power-law
continuum model. (b) best-fit residuals which reveal the presence of a
broad feature centered near 0.9~keV, together with narrow absorption
features at 6.65 and 6.95~keV. (c) residuals when an edge at 0.98~keV
and two absorption Gaussian lines at 6.65 and 6.95~keV are included in
the spectral model. (d) deconvolved photon spectrum and the best-fit
model including an edge at 0.98~keV and two absorption lines at 6.65
and 6.95~keV. The individual components are indicated. The best-fit
parameters for this model are given in Table~\ref{tab:spectrum}. The
other absorption edges visible in the spectrum are attributed to
neutral material in the interstellar medium.}
\label{fig:spectrum}
\end{figure*}

\subsection{EPIC pn spectra}

\subsubsection{Persistent emission}

Intervals flagged with thick horizontal lines in Fig.~\ref{fig_pn_lc}
were selected as the persistent emission for the EPIC pn spectral
analysis. They correspond to phases in the range 0.25--0.92. The
interval corresponding to the intermediate dipping activity visible
near 5.7~h has been excluded. The resulting persistent emission pn
spectrum has an exposure of 9.9~ks.

An absorbed disk-blackbody plus power-law model fits the overall
continuum reasonably well. The reduced \chisq\ (\rchisq) is 1.48 for
229 degrees of freedom (d.o.f.). The absorption, \nh, is $(0.53
\pm 0.01)$~\ttnh, the blackbody effective temperature at the innermost
radius, $kT_{\rm in}$, is 3.2~keV and \phind\ is 2.9. We note that the
disk-blackbody component dominates the spectrum above $\sim$3~keV.  It
is probably over-estimated because of the lack of data above 10~keV.
The inferred inner radius of the disk is remarkably small
(0.5~km assuming an inclination of 70\degree) and thus without any
physical meaning. We caution that the power-law and disk-blackbody
continuum model used here is not unique and other combinations of
simple models can also provide acceptable fits.  However, we focus
here on the discovery of absorption features whose properties depend
only weakly on the chosen continuum model. Detailed
modelling of the continuum emission and discussion of its physical
interpretation are the subject of a separate paper (Webb et al., in
preparation). The power-law and disk-blackbody model allows the
underlying continuum to be reliably modelled so that the absorption
feature properties can be derived. Furthermore, this model allows the
spectral changes between the persistent and the dipping emission to be
modelled in a coherent way, within the framework of the complex
continuum approach (see Sect.~\ref{sec:pndipping}).

The observed count spectrum fit with the absorbed disk-blackbody
(3.2~keV) and power-law model (\phind\ of 2.9) is shown in
Fig.~\ref{fig:spectrum}a.  Examination of the residuals from this fit
(Fig.~\ref{fig:spectrum}b) reveals a structure around 0.9~keV and two
deep negative structures around 7~keV.  The smaller structures at
$\sim$1.8~keV and $\sim$2.2~keV are probably due to an incorrect
instrumental modelling of the Si CCD and Au mirror edges.

In order to model the structure around 0.9~keV, a Gaussian emission
feature with an energy of $0.87^{+0.02}_{-0.03}$~keV and width
($\sigma$) of $<$74~eV was first added to the previous model. This
improves the fit (\rchisq\ of 1.35 for 227 d.o.f.).  However, an even
better fit is obtained if the Gaussian is replaced by an absorption
edge at $0.98 \pm 0.02$~keV with an optical depth, $\tau$, of $0.10 \pm
0.02$. This results in a \rchisq\ of 1.30 for 226 d.o.f. An F-test
indicates that the probability for such an improvement occuring by
chance is $1.7 \times 10^{-7}$.  Thus, including an edge significantly
improves the fit.  We note that the RGS data support the edge
interpretation independently using a different continuum model (see
Sect.~\ref{sec:rgs}).  The feature could be a K absorption edge from
moderately ionized Ne ions, or an L absorption edge from moderately
ionized Fe ions, or an edge complex from both kind of ions, since
these astrophysically abundant elements have a series of
photo-ionization thresholds around 0.9~keV (the photo-ionization
thresholds of \ion{Ne}{v} K and of \ion{Fe}{x} L being 0.987 and
0.959~keV, respectively).

\begin{table}
\begin{center}
\caption[]{Best-fit to the 0.6--10~keV \nineteen\ pn persistent
emission spectrum using a disk-blackbody model with an inner
temperature \ktin\ and normalization $k_{\rm DBB}$ given by $(R_{in}/d)^2{\bf \cos{i}}$ where $R_{in}$ in the inner radius in km, $d$
the distance in units of 10~kpc and $i$ the inclination and a
power-law with a photon index, $\alpha$ and a normalization at 1~keV,
$k_{\rm PL}$, in unit of
\kevnorm, together with an edge with an energy, \eedge, and an optical
depth, $\tau$, and two Gaussian absorption lines with their energy,
\eline, width, \sig\ and equivalent width, \ew.}
\begin{tabular}{llc}

\hline
\hline
\noalign {\smallskip}
\multicolumn{3}{c}{EPIC pn persistent emission}\\
\noalign {\smallskip}
\hline
\noalign {\smallskip}
Component & Parameter 	& Value          \\ 
\noalign {\smallskip}
\hline
\noalign {\smallskip}

Disk-blackbody	& \ktin\ (keV)	& 3.18 $\pm$ 0.08 \\
		& $k_{\rm DBB}$ & 0.106 $^{+0.012}_{-0.006}$ \\
\noalign {\smallskip}
\noalign {\smallskip}

Power-law	& \phind 	& 3.20 $\pm$ 0.09 		\\
		& $k_{\rm PL}$	& 0.113 $\pm$ 0.006 		\\
\noalign {\smallskip}
\noalign {\smallskip}

Edge		& \eedge\ (keV)	& 0.98 $\pm$ 0.02	\\
		& $\tau$	& 0.11 $\pm$ 0.03	\\

\noalign {\smallskip}
\noalign {\smallskip}

\fetfive\ \ka 	& \eline\ (keV) & 6.65 $^{+0.05}_{-0.02}$ \\
absorption	& \sig\ (eV)	& $<$100	\\
Gaussian	& \ew\ (eV)		& $-30~^{+8}_{-12}$ \\
\noalign {\smallskip}
\noalign {\smallskip}

\fetsix\ \ka 	& \eline\ (keV) & 6.95 $^{+0.05}_{-0.04}$ \\
absorption	& \sig\ (eV)	& $<$140 		\\
Gaussian	& \ew\ (eV)	& $-30~^{+11}_{-12}$     \\

\noalign {\smallskip}
\noalign {\smallskip}
\noalign {\smallskip}
	\multicolumn{2}{c}{\nh\ (atom~cm$^{-2}$)}	 & $(0.59 \pm 0.02) \times 10^{22}$ \\
\noalign {\smallskip}
	\multicolumn{2}{c}{\rchisq\ (d.o.f.)} & 0.99 (221) \\
\noalign {\smallskip}
	\multicolumn{2}{c}{$L_{\rm 0.6-10~keV}$ (\ergsec, at 9.3~kpc)} 	& $ 4.5 \times 10^{36}$ \\
\noalign {\smallskip}
\noalign {\smallskip}
\hline
\label{tab:spectrum}
\end{tabular}
\end{center}
\end{table}

\begin{figure}[h!]
\includegraphics[width=0.5\textwidth]{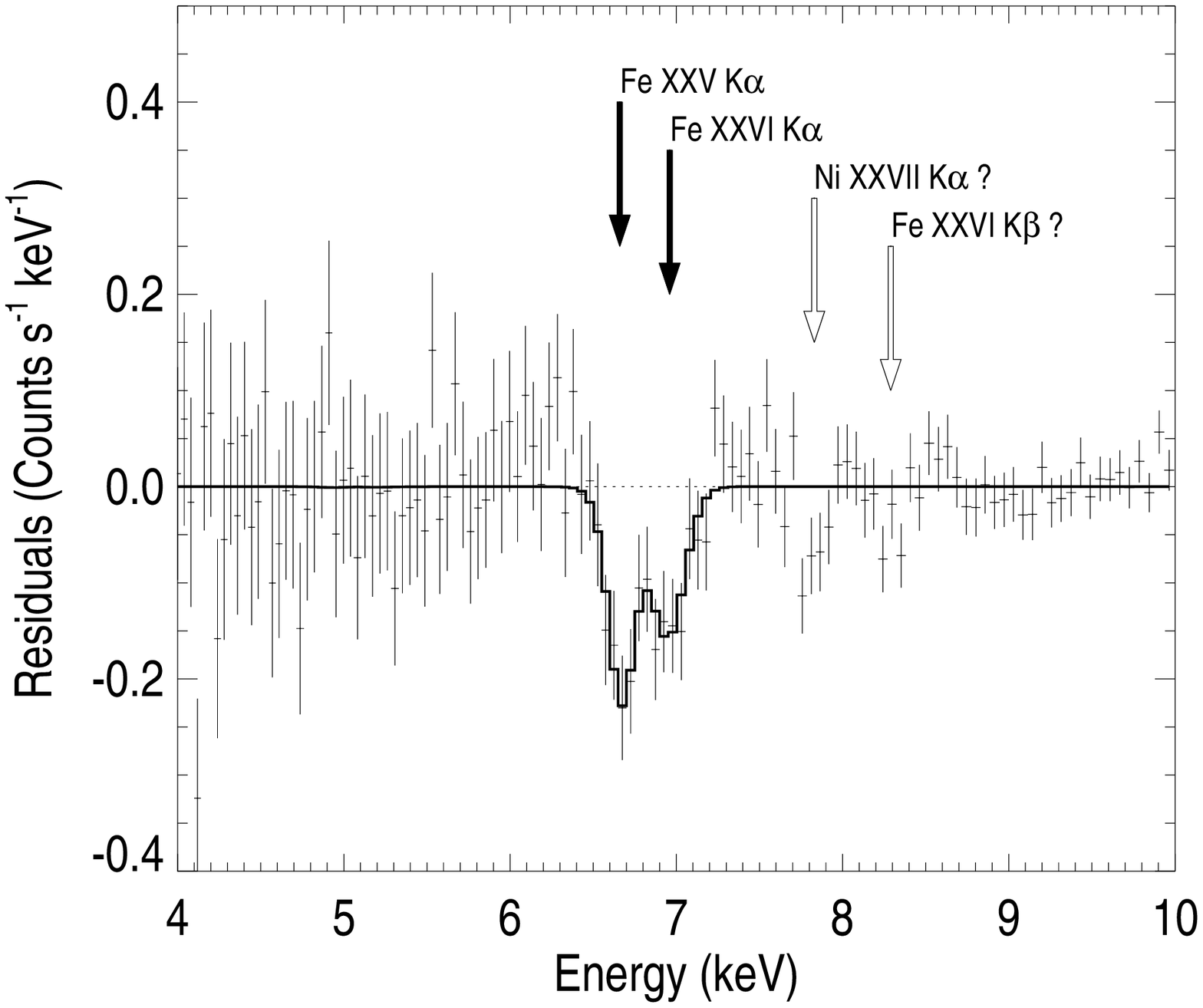}
\caption{Residuals in the 4--10~keV energy range when the best-fit
model given in Table~\ref{tab:spectrum} is fit to the pn spectrum of
the \nineteen\ persistent emission. The normalisations of the narrow
absorption features at 6.65~keV and 6.95~keV have been set to
zero. Theses two absorption features, identified with \fetfive\ \ka\
and \fetsix\ \ka, are shown with filled arrows. Fainter absorption
features shown with non-filled arrows at 7.83 and 8.29~keV may be
present and have energies consistent with \nitseven\ \ka\ and \fetsix\
\kb\ transitions.}
\label{fig:zoom_residuals}
\end{figure}

The two deep negative residuals near 7~keV were modelled using two
narrow Gaussian absorption lines centered on 6.65~keV and
6.95~keV. This results in a \rchisq\ of 0.99 for 221 d.o.f.  An F-test
indicates that the probability for such an improvement occuring by
chance is $2.0 \times 10^{-12}$. The measured energy of $6.65
^{+0.05}_{-0.02}$~keV of the first feature is consistent with that of
\fetfive\ \ka, while the measured energy of $6.95
^{+0.05}_{-0.04}$~keV of the second feature is consistent with that of
\fetsix\ \ka\ \citep[see][ Table 5, for a list of absorption line
candidates and references]{1915:kotani00apj}. Thus, we identify the
observed absorption features with these resonant transitions from
highly ionized Fe ions (He-like and H-like ions, respectively). 

Table~\ref{tab:spectrum} gives the best-fit parameters of the model
consisting of an edge (0.98~keV), a disk-blackbody (3.2~keV), a
power-law (\phind\ of 3.2) and two narrow absorption Gaussian lines
(6.65 and 6.95~keV). This model and its components are shown together
with the deconvolved spectrum in Fig.~\ref{fig:spectrum}d. The
residuals from this fit are shown in Fig.~\ref{fig:spectrum}c. An
expanded view of the residuals in the 4--10~keV range is shown in
Fig.~\ref{fig:zoom_residuals}.

In addition to the clearly detected absorption features at 6.65~keV
and 6.95~keV (Fig.~\ref{fig:zoom_residuals}, filled arrows), there is
evidence for fainter absorption features at 2.67~keV, 7.82~keV and
8.29~keV (Fig.~\ref{fig:spectrum}c and Fig.~\ref{fig:zoom_residuals}).
When absorption Gaussian lines are included, their derived energies
are $2.67 ^{+0.04}_{-0.07}$~keV, $7.82^{+0.03}_{-0.07}$~keV and
$8.29^{+0.08}_{-0.12}$~keV, and their derived \ews\ are $-6 \pm 3$~eV,
$-21 ^{+9}_{-12}$~eV and $-19 ^{+11}_{-12}$~eV, respectively. Although
these features are marginal, it is interesting to note that the
measured energies are consistent with that of \ssixteen, \nitseven\
\ka\ and \fetsix\ \kb, respectively. \nitseven\
\ka\ and \fetsix\ \kb\ features have been seen in the ASCA spectrum of
\grs\ \citep{1915:kotani00apj}, and there was evidence for their
presence in the XMM-Newton spectrum of \bigdip\
\citep{1624:parmar02aa} too. A P~Cygni \ssixteen\ feature is reported in 
\cir\ \citep{cirx1:brandt00apjl}. Thus, we consider the marginal detections
of absorption features at 2.67, 7.82 and 8.29~keV in the spectrum of
\nineteen\ as evidence for the possible presence of \ssixteen, \nitseven\ 
\ka\ and \fetsix\ \kb\ absorption lines.

We have searched for K absorption edges in the pn persistent emission
spectrum due to neutral Fe, \fetfive\ and \fetsix\ expected at 7.12,
8.76 and 9.28~keV. The upper-limits to the optical depth of an edge at
7.12~keV, 8.76~keV and 9.28~keV are 0.05, 0.06 and 0.02, respectively.

As a broad emission feature at $\sim$6 keV interpreted as fluorescent
line emission from neutral Fe was detected in several spectra from
\nineteen, we tried to include one in the model. Including a Gaussian
emission feature with an energy of $6.6^{+0.2}_{-0.6}$~keV and width
$450 \pm 350$~eV to the model described in Table~\ref{tab:spectrum}
results in a \rchisq\ of 0.96 for 218 d.o.f. An F-test indicates that
the probability for such an improvement occuring by chance is
0.035. Thus, the fit is only marginally improved when the emission
line is included.  Therefore, the emission line was not kept in the
model. We note however that the significance of such an emission line
depends on the model chosen for the underlying continuum.  If the high
energy part of the spectrum is modelled using a cut-off power-law
instead of a disk-blackbody as in the previous model, then the
inclusion of a Gaussian emission line at $\sim$6~keV significantly
improves the fit.  For instance, a good fit (\rchisq\ of 0.99 for 217
d.o.f.) of the spectrum is obtained using a model consisting of a
disk-blackbody with \ktin\ of $\sim$0.3~keV, a cut-off power-law with
\phind\ of $\sim$1.6 and \ecut\ of $\sim$15~keV, in addition to an 
edge near 0.98~keV and two absorption lines near 6.65 and 6.95~keV
with similar properties as in Table~\ref{tab:spectrum}, and in
addition to a Gaussian emission line with an energy of $6.2 \pm
0.3$~keV and width of $980^{+450}_{-280}$~eV. An F-test indicates that
the probability of improving the fit by chance by adding the latter
line is $3 \times 10^{-7}$. So, the Gaussian emission line is highly
significant in that case. This difference illustrates rather well the
difficulties of spectral fitting with moderate spectral resolution CCD
data. The presence of the emission line is further discussed in the
separate paper devoted to the continuum modelling.

\begin{figure}[t!]
\includegraphics[width=0.47\textwidth]{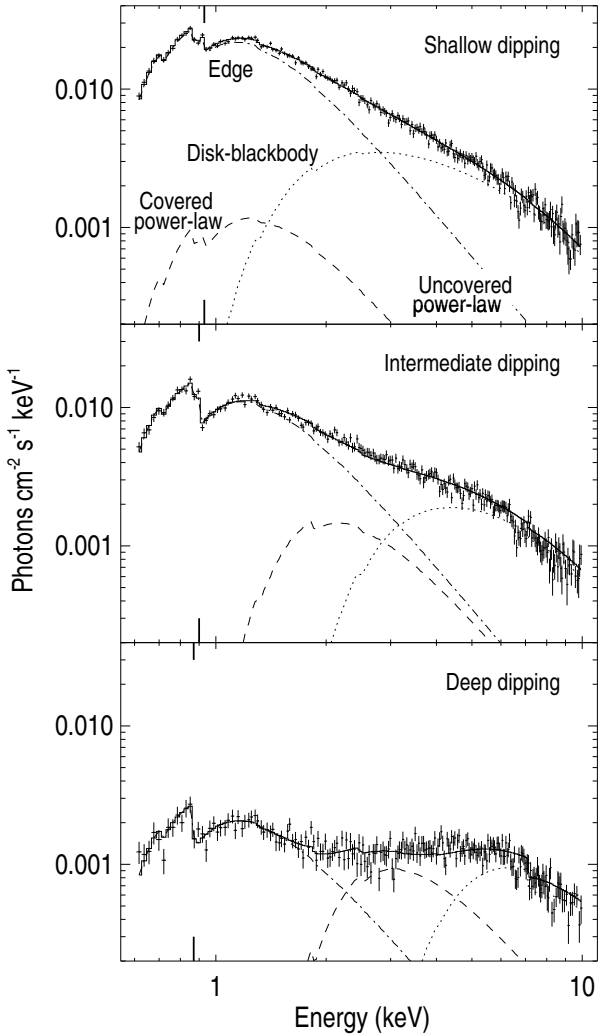}\\
\vspace{-0.7cm}
\caption{Deconvolved pn intensity-selected dipping spectra of
\nineteen\ fit using a complex continuum model. The different components 
of the model are indicated in the upper panel: the edge, the
disk-blackbody (dotted line), the power-law partly covered (dashed
line) and partly uncovered (dashed-dotted line). Vertical tick marks
indicate the position of the edge in each spectrum. The decrease of
the edge energy as \nineteen\ evolves from shallow to deep dipping is
clearly visible.}
\label{fig:dipspectra}
\end{figure}

\begin{figure}[ht!]
\includegraphics[width=0.47\textwidth]{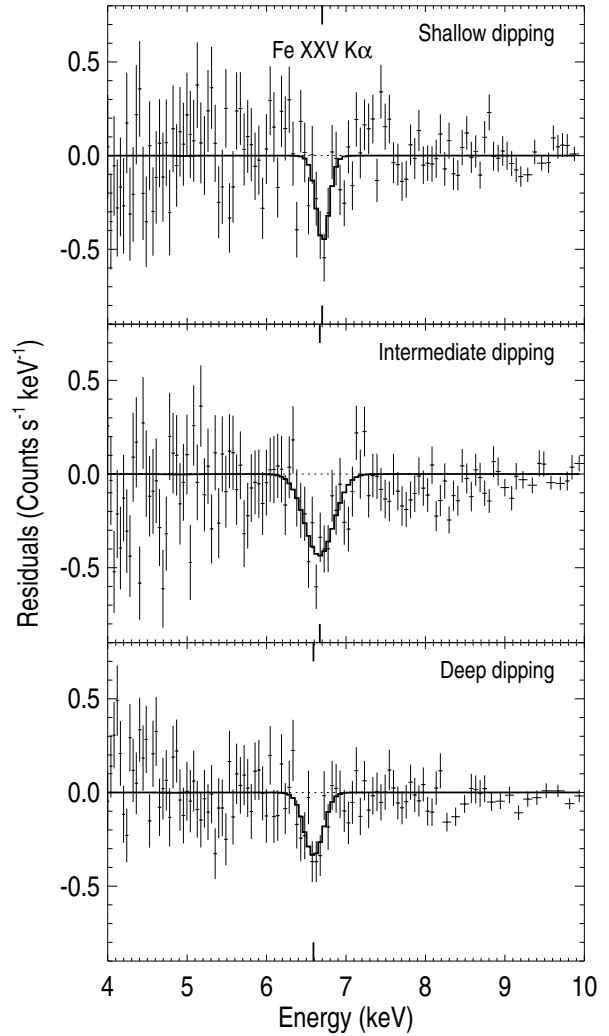}\\
\vspace{-0.7cm}
\caption{4--10~keV residuals when the best-fit
complex continuum model and absorption lines are fit to the pn
intensity-selected \nineteen\ dipping spectra. The normalizations of
the narrow absorption features have been set to 0. An absorption
feature near 6.7~keV, identified with \fetfive\ \ka, is detected in
each spectrum and its energy is indicated by vertical tick marks.}
\label{fig:dipresiduals}
\end{figure}

\subsubsection{Dipping emission}
\label{sec:pndipping}

Intervals with phases in the range 0.92--0.25 were used to investigate
the dipping emission.  Within these dipping intervals, three spectra
were extracted based on intensity selection criteria.  Events
corresponding to a background-subtracted pn 0.6--10~keV count rate in
the range 40--60, 20--40 and 0--20~s$^{-1}$ (see Fig.~\ref{fig_pn_lc})
were extracted to form ``shallow'', ``intermediate'', and ``deep''
dipping spectra, respectively.

To quantify the spectral evolution from persistent to deep dipping
emission and extract the properties of lines during dipping, the
complex continuum approach was used. Discussion on other possible
continuum models is reserved to a separate paper.  Our reference model
for the complex continuum approach is the best-fit to the persistent
spectrum using a model including the edge, the disk-blackbody and the
power-law components. The best-fit \nh\ (applied to both components)
is $(0.58 \pm 0.02)$~\ttnh, \ktin\ is $3.13 ^{+0.09} _{-0.08}$~keV,
\phind\ is $3.17 \pm 0.09$, \eedge\ is $0.98 \pm 0.02$~keV and $\tau$
is $0.10 \pm 0.03$ (see also Table~\ref{tab:complexcontinuum}).
We applied the complex continuum model in the following form:
Edge*[AB$^{\rm DBB}$*DBB~+~AB$^{\rm Gal}$*[AB$^{\rm PL}$*$f$ +
(1-$f$)]*PL], where DBB represents the disk-blackbody component
(point-source), and PL the power-law component (extended).  $f$ is the
covering fraction ($0\lid f \lid 1$). The AB terms represent
absorption of the form e$^{-\sigma N_{\rm H}}$ where $\sigma$ are the
photoelectric cross-sections and $N_{\rm H}$ the column density.
AB$^{\rm Gal}$ represents Galactic absorption. AB$^{\rm PL}$
reprensents the absorption applied to the covered fraction of the
power-law emission, in addition to the Galactic absorption. AB$^{\rm
DBB}$ represents the absorption applied to the disk-blackbody
component and includes Galactic absorption.

\begin{table}[!b]
\begin{center}
\caption[]{Spectral parameters of the \fetfive\ \ka\ 
absorption line detected during
intensity-selected dipping intervals. The feature is
modelled as a Gaussian line added to the disk-blackbody component of
the complex continuum model. \eline, $\sigma$ and \ew\ are the
line energy, width and equivalent width, respectively.}
\begin{tabular}{llccc}
\hline
\hline
\noalign {\smallskip}
 \multicolumn{5}{c}{EPIC pn dipping emission}\\
\noalign {\smallskip}
\hline
\noalign {\smallskip}
 & & Shallow & Intermediate & Deep \\
\noalign {\smallskip}
\hline
\noalign {\smallskip}
\multicolumn{5}{l}{\fetfive\ \ka\ absorption Gaussian}\\
\noalign {\smallskip}
&\eline\ (keV) &  6.70 $\pm$ 0.05 & 6.67 $^{+0.06}_{-0.05}$ & 6.59 $\pm$ 0.05\\
&\sig (eV) & $<$110 & 160 $^{+80}_{-60}$ & $<$173\\
&$EW$ (eV) & -74 $^{+20}_{-27}$ & -168 $^{+44}_{-46}$ & -119 $^{+45}_{-50}$\\
\noalign {\smallskip}
\hline
\label{tab:dipping}
\end{tabular}
\end{center}
\end{table}

\begin{figure}[ht!]
\includegraphics[width=0.49\textwidth]{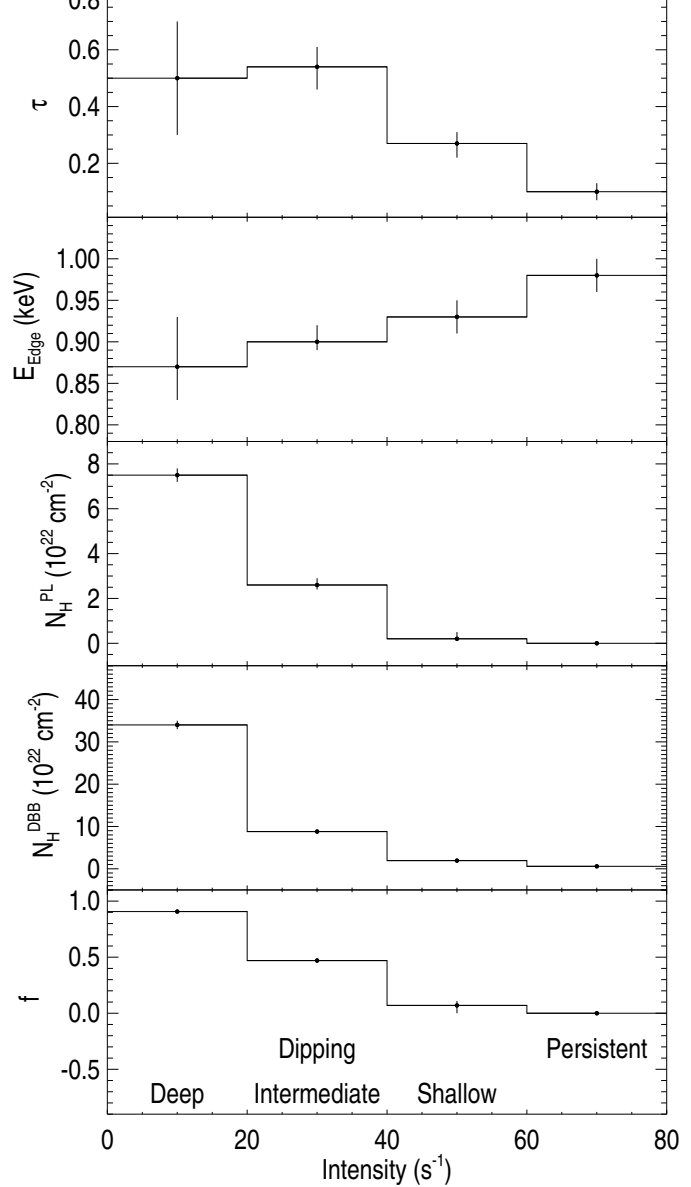}\\
\caption{Best-fit parameters to the intensity-selected dipping spectra
using the complex continuum approach, plotted as a function of
intensity (see text and Table~\ref{tab:complexcontinuum}). The best-fit
parameters to the persistent spectrum using the basis model for the
complex continuum approach are also shown.}
\label{fig:parameters}
\end{figure}

\nhgal\ was fixed to 0.58~\ttnh, the value obtained from the fit to
the persistent spectrum. When $f$ is set to 0, i.e. when the power-law
component is uncovered, the complex continuum model is equivalent to
our reference model applied to the persistent emission, as expected.
To fit each intensity-selected dipping spectrum, only $f$, \nhdbb, and
\nhpl\ are allowed to vary, whereas the parameters and normalizations
of the disk-blackbody and power-law components are fixed to their
persistent emission values.  The parameters of the edge (energy and
optical depth) were also fixed to the persistent emission values. This
gave reasonably good fit to each dipping spectrum, with \rchisq\
values of 1.28 (227 d.o.f.), 2.03 (225 d.o.f.) and 1.30 (217 d.o.f.) 
for the shallow, intermediate and deep dipping spectra,
respectively. 
However, inspection of the residuals from this fits
reveals broad structures below $\sim$1~keV. These structures are well
accounted for if the edge parameters are allowed to vary, instead of
being fixed to the persistent emission values. When the edge
parameters are allowed to vary, the \rchisq\ values become 1.14 (225
d.o.f.), 1.57 (223 d.o.f.)  and 1.26 (215 d.o.f.) for the shallow,
intermediate and deep dipping spectra, respectively. F-tests indicate
that the probability for such an improvement occuring by chance is
$8.2 \times 10^{-7}$, $1.1 \times 10^{-13}$ and $6.6 \times 10^{-2}$,
respectively. Thus, while the fit of the deep dipping spectrum is only
marginally improved, the fit of the shallow and intermediate spectra
are significantly improved when the edge parameters are allowed to
vary. The intensity-selected dipping spectra fit in such a way are
shown in Fig.~\ref{fig:dipspectra}. The corresponding best-fit
parameters are given in Table~\ref{tab:complexcontinuum} and plotted
as a function of intensity in Fig.~\ref{fig:parameters}.  
The persistent emission parameters are also shown for comparison.
$f$,
\nhpl\ and \nhdbb\ are all observed to increase as \nineteen\ evolves
from its persistent to its deep dipping state.  The power-law emission
is progressively covered until $f$ reaches 91\% and \nhpl\ 7.5~\ttnh\
in the deepest dipping spectrum. The disk-blackbody emission is
strongly absorbed, with \nhdbb\ reaching 34~\ttnh\ in the deepest
dipping spectrum.
Thus, dipping is well accounted for by large
increases of column density for the point-like disk-blackbody
emission, combined with the extended power-law emission being
progressively covered by the absorber. This is consistent with
previous results of the complex continuum approach applied to
\nineteen\ and other dipping sources \citep[e.g.,
][]{1916:church97apj}. An interesting new result is the decrease of
the edge energy as \nineteen\ evolves from its persistent towards its
deepest dipping state. Assuming the edge actually represents a complex
of edges from ions in a range of ionization states, the decrease of
energy is consistent with a decrease in the average ionization
level. This suggests that during dipping, the additional absorbing
material is less ionized than during the persistent emission. The
increase of the edge optical depth from the persistent to the deepest
dipping state is consistent with more absorbing material being present
in the line of sight during dipping.

\begin{table*}[!th]
\begin{center}
\caption[]{Best-fits to the intensity-selected dipping spectra using
the complex continuum model (see text). The best-fit to the persistent
spectrum used as a basis for the complex continuum approach is also
given for comparison.}
\begin{tabular}{lccccccc}
\hline
\hline
\noalign {\smallskip}
EPIC pn  & \nhdbb & \nhpl & $f$ & \eedge & $\tau$ & \rchisq & d.o.f. \\
	 & ($10^{22}$ cm$^{-2}$)      & ($10^{22}$ cm$^{-2}$) & & (keV)& & &\\ 
\hline
\noalign {\smallskip}
Persistent emission	& \multicolumn{2}{c}{\nh = (0.58 $\pm$ 0.02)} & (0) & 0.98 $\pm$ 0.02& 0.10 $\pm$ 0.03 & 1.30 & 227 \\
\noalign {\smallskip}
Shallow dipping	& 1.93 $^{+0.12}_{-0.07}$ & 0.2 $^{+0.3} _{-0.1}$ & 0.07 $^{+0.04}_{-0.07}$ & 0.93 $\pm$ 0.02 & 0.27 $^{+0.04}_{-0.05}$ & 1.14 & 225 \\
Intermediate dipping & 8.8 $\pm$ 0.4 & 2.6 $^{+0.3}_{-0.2}$ & 0.47 $\pm$ 0.02 & 0.90 $^{+0.02}_{-0.01}$ & 0.54 $^{+0.07}_{-0.08}$ & 1.57 & 223 \\
Deep dipping 	& 34 $\pm$ 1 & 7.5 $\pm$ 0.3 & 0.907 $\pm$ 0.006 & 0.87 $^{+0.06}_{-0.04}$ & 0.5 $\pm$ 0.2 & 1.26 & 215 \\
\noalign {\smallskip}
\hline
\label{tab:complexcontinuum}
\end{tabular}
\end{center}
\end{table*}

Examination of the intensity-selected dipping spectra
(Fig.~\ref{fig:dipspectra}) reveals narrow absorption features near
6.7~keV. Adding a Gaussian line in absorption at $\sim$6.7~keV to the
complex continuum model gives \rchisq\ of 0.99 (222 d.o.f.), 1.27 (220
d.o.f.) and 1.12 (212 d.o.f.)  for the shallow, intermediate and deep
dipping spectra, respectively. F-tests indicate that the probability
for such an improvement occuring by chance is $3.5 \times 10^{-7}$,
$9.2 \times 10^{-11}$ and $4.3 \times 10^{-6}$, respectively. The
properties of the lines are given in Table~\ref{tab:dipping}, and an
expanded view of the residuals to the intensity-selected dipping
spectra is shown in Fig.~\ref{fig:dipresiduals}.  The measured
energies of the absorption lines are all around 6.7~keV and thus
consistent with that of \fetfive\ \ka, already detected in the
persistent spectrum. However, there is some evidence for a slight
decrease of the Gaussian energy as \nineteen\ evolves from shallow to
deep dipping. Considering that the observed absorption feature, with
limited statistics and resolution, may not be only due to one ion, but
may contain unresolved contributions from other ions, this trend,
although marginal, is consistent with the absorbing material becoming
less ionized from shallow to deep dipping. The absorption feature
detected during the deep dipping spectrum could be due to the
combination of \ka\ absorption lines from \fetfive,
\fetfour\ and less ionized ions, with a proportion of \fetfour\ and
less ionized ions larger than during the intermediate dipping.


In addition to the clearly detected \fetfive\ \ka\ absorption line,
there is evidence for an absorption feature around 8.3~keV in the deep
dipping spectrum (Fig.~\ref{fig:dipresiduals}). When an absorption
line is included, its measured energy is $8.35 \pm 0.05$~keV. This
energy is close to but not consistent with that of \fetfive\
\kb. There is no obvious feature expected to be strong at that
energy. If the feature is modelled as an edge rather than a Gaussian
line, the best-fit edge energy is $8.0 ^{+0.3} _{-0.4}$~keV and $\tau
= 0.3 \pm 0.1$. A likely explanation for this marginal feature is a
combination of K absorption edges from moderately ionized Fe
(\ion{Fe}{xii} and higher) and neutral or slightly ionized Ni, with
possibly the contribution of absorption lines as well.

We set upper-limits of $-54$~eV, $-72$~eV and $-50$~eV on the \ew\ of
a narrow \fetsix\ \ka\ in shallow, intermediate and deep dipping
spectra, respectively (by including a Gaussian with an energy fixed to
6.97~keV, the rest energy of the transition, and a width fixed to
0). The marginally significant lines seen in the persistent spectrum
are not evident in the dipping spectra.

\subsection{RGS spectra}
\label{sec:rgs}

The dipping activity is clearly visible in the RGS light curve. The
same intervals as used for the pn were used to extract persistent and
dipping RGS spectra.  We did not further divide the dipping intervals
using intensity criteria as was done for pn, because of the low RGS
count rate during the dips.  RGS1 and RGS2 spectra were fit together
with a unique model to better constrain spectral modelling, but the
normalizations were allowed to vary independently to account for
calibration differences.  The ranges 0.903--1.177 and 0.516--0.602~keV
are not present in the data from RGS1 and RGS2, respectively, because
of malfunction of the drive electronics for one CCD chip in each
spectrometer.

\begin{figure}[ht!]
\includegraphics[width=0.5\textwidth]{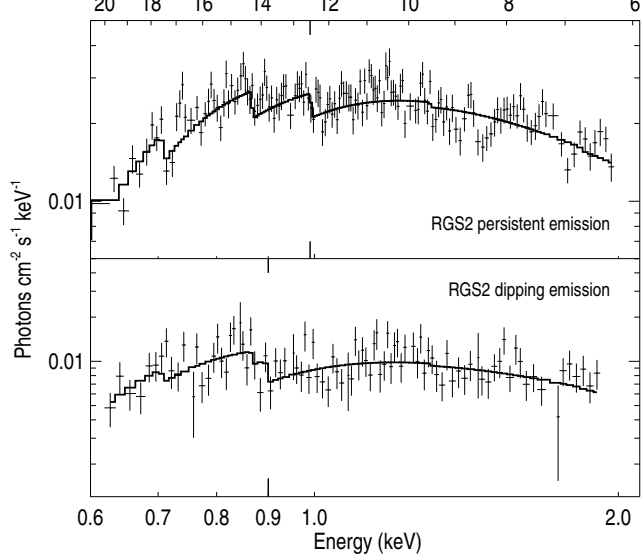}\\
\caption{Deconvolved RGS2 spectra of the \nineteen\ persistent (top
panel) and dipping emission (bottom panel) fit using a power-law and
an edge model (thick line). The best-fit models were obtained by
fitting RGS1 and RGS2 spectra simultaneously (see
Table~\ref{tab:rgs}), but only RGS2 spectra are shown for
clarity. Vertical tick marks indicate the position of the edge
included in each spectrum.}
\label{fig:rgs}
\end{figure}

The persistent emission RGS spectrum was first modelled using an
absorbed power-law.  This results in a moderately good fit with a
\rchisq\ of 1.26 for 891 d.o.f. However, examination of the residuals
reveals a structure around 0.9~keV similar to the one observed in the
pn spectrum and well accounted for by an edge
(Fig.~\ref{fig:spectrum}b and Table~\ref{tab:spectrum}). Therefore, we
included such an edge to the previous model of the RGS spectrum. This
gives a \rchisq\ of 1.22 for 889 d.o.f. An F-test indicates that the
probability for such an improvement occuring by chance is $4.8 \times
10^{-7}$. Thus, the fit is significantly improved when the edge is
included.  The RGS persistent emission spectrum and the best-fit model
using a power-law and an edge are shown in Fig.~\ref{fig:rgs} (top
panel), and the best-fit parameters are given in
Table~\ref{tab:rgs}. The edge energy of $0.99 \pm 0.02$~keV measured
from the RGS spectrum is fully consistent with that obtained from the
EPIC pn spectrum.

A good fit to the dipping emission RGS spectrum is obtained using an
absorbed power-law model (\rchisq\ of 1.06 for 171 d.o.f.). As an edge
around 0.9~keV was detected in the dipping emission pn spectra, we
included an edge to the previous model of the RGS spectrum. This gives
a \rchisq\ of 1.00 for 169 d.o.f. An F-test indicates that the
probability for such an improvement occuring by chance is 0.0027,
indicating that the edge is detected at a 99.7\% confidence level
(3\sig). The RGS dipping emission spectrum and the corresponding
best-fit model using a power-law and an edge are shown in
Fig.~\ref{fig:rgs} (bottom panel), and the best-fit parameters are
given in Table~\ref{tab:rgs}. The measured edge energy of $0.90
^{+0.07}_{-0.03}$~keV obtained from the dipping RGS spectrum is
consistent with the mean value obtained from the three
intensity-selected dipping pn spectra
(Table~\ref{tab:complexcontinuum}). Thus, RGS results strengthen the
conclusion drawn from pn results that the edge energy is less during
dipping emission than during persistent emission, which is consistent
with the absorbing material being less ionized during dipping.

We note that the column densities obtained from the RGS persistent and
dipping spectra (\nh\ of 0.41~\ttnh\ and 0.51~\ttnh, respectively) are
roughly equal to each other and roughly consistent with that obtained
from EPIC pn spectra. Indeed, the low-energy part of pn spectra (where
RGS is sensitive) is dominated by the uncovered and unabsorbed
component (here the power-law) of the complex continuum model (see
Fig.~\ref{fig:dipspectra}). This component is only affected by the
Galactic absorption, assumed to be 0.58~\ttnh\ within the complex
continuum model. When \nh\ is forced to be $\ge$0.58~\ttnh, almost as
good fits as those presented in Table~\ref{tab:rgs} are obtained to
the RGS spectra, with best-fit \nh\ of 0.58~\ttnh\ in both cases,
similar parameters for the edge and \phind\ of $\sim$2.7 and $\sim$2.9
in the persistent and dipping spectrum, respectively. No acceptable
fit is found to the RGS dipping spectrum by fixing \nh\ to an higher
value (e.g. 2~\ttnh), since such an absorption implies basically no
counts below 1~keV, whereas counts below 1~keV are actually present.

An absorption feature is evident near 1.5~keV (8.5~\ang) in the RGS2
persistent emission spectrum shown in Fig.~\ref{fig:rgs}. However this
feature is less obvious in RGS1. The inclusion of a Gaussian
absorption line with an energy of $1.48 \pm 0.01$~keV, a width of
$<$41 ~eV, and an \ew\ of $-7 \pm 3$~eV in the combined RGS spectrum
is marginally significant. However, it is interesting to note that the
measured energy is consistent with that of \ion{Mg}{xii} \ka\ at
1.47~keV.  Thus, we consider that there is marginal evidence for the
presence of an absorption line from highly ionized Mg (H-like) in
\nineteen. The upper-limit on the \ew\ of a narrow \ion{Mg}{xii} \ka\ 
absorption line in the pn persistent spectrum is $-4.5$~eV.  The
upper-limit on the \ew\ of a narrow \ion{Mg}{xii} \ka\ absorption line
in the RGS dipping spectrum is $-18$~eV.  The upper-limits on a narrow
\ion{Ne}{x} absorption line at 1.02~keV (12.19~\ang), a feature
detected e.g. in
\mxb\ \citep{1658:sidoli01aa}, are $-5$ and $-9$~eV in the RGS persistent
and dipping spectra, respectively.

\begin{table}[!t]
\begin{center}
\caption[]{Best-fit to RGS persistent and dipping emission spectra of
\nineteen\ using a power-law model with a photon index, \phind,
together with an edge with an energy, \eedge, an optical depth,
$\tau$.  RGS1 and RGS2 spectra are fit together, but the RGS1 and RGS2
power-law normalizations at 1~keV, $k_{\rm RGS1}$ and $k_{\rm RGS2}$
(\kevnorm), are allowed to vary independently.}
\begin{tabular}{lllcc}

\hline
\hline
\noalign {\smallskip}

\multicolumn{3}{l}{RGS1 \& RGS2} &  Persistent & Dipping\\
\noalign {\smallskip}
\hline
\noalign {\smallskip}

&Power-law	& \phind 	& 2.4 $^{+0.2}_{-0.1}$	& 2.1 $^{+0.4}_{-0.3}$\\
&		& $k_{\rm RGS1}$	& 0.101 $^{+0.01}_{-0.009}$ & 0.036 $^{+0.01}_{-0.007}$\\
&		& $k_{\rm RGS2}$	& 0.092 $^{+0.008}_{-0.009}$	& 0.031 $^{+0.008}_{-0.006}$\\
\noalign {\smallskip}
\noalign {\smallskip}

&Edge		& \eedge\ (keV)	& 0.99 $\pm$ 0.02	& 0.90 $^{+0.07}_{-0.03}$\\
&		& $\tau$	& 0.22 $\pm$ 0.06	& 0.3 $^{+0.2}_{-0.1}$\\

\noalign {\smallskip}
\noalign {\smallskip}
&	\multicolumn{2}{c}{\nh\ ($10^{22}$ cm$^{-2}$)}	 & $0.51 \pm 0.03$ & $0.42 \pm 0.07$\\
\noalign {\smallskip}
&	\multicolumn{2}{c}{\rchisq\ (d.o.f.)} & 1.22 (889) & 1.00 (169)\\

\noalign {\smallskip}
\hline

\label{tab:rgs}
\end{tabular}
\end{center}
\end{table}

\section{Discussion}

We have modelled the EPIC pn continuum of \nineteen\ during persistent
emission intervals using a combination of disk-blackbody and power-law
models together with an absorption edge at 0.98~keV and narrow
absorption lines at 6.65 and 6.95~keV. We have characterized the
spectral changes in the EPIC pn data during dipping intervals
following the complex continuum approach, and using the same continuum
components in order to investigate the properties of the narrow
absorption features. Dipping is well accounted for by large increases
in \nh\ of the point-like disk-blackbody emission, combined with the
extended power-law emission being progressively covered by the
absorber. The edge energy decreases from 0.98~keV to 0.87~keV as
\nineteen\ evolves from persistent to deep dipping. This result is
consistent with that obtained from RGS data. This suggests that the
detected edge represents a complex of edges from moderately ionized Ne
and/or Fe, and that the average ionization level decreases from
persistent to deep dipping.  This implies that during dipping, the
absorbing material in the line of sight is less ionized than during
the persistent emission. Photo-ionization is likely to be the dominant
ionization process in X-ray binaries because of the high radiation
field.  Since the photo-ionization parameter, $\xi = L /n_{\rm e} \,
r^{2}$ \citep[e.g., ][]{kallman82apjs}, where $L$ is the luminosity of
the ionizing source, $n{\rm _e}$ the electron density and $r$ the
distance to the ionizing source, depends on $r^{-2}$ and $n{\rm
_e}^{-1}$, the decrease of the average ionization level during dipping
indicates that the additional absorbing material during dipping is
either more distant from the ionizing source and/or denser than the
absorbing material present during persistent emission intervals. This
supports the model where the dips are due to relatively cold clouds
near the outer edge or circularization radius, of the disk.
\begin{table*}[!ht]
\begin{center}
\caption[]{X-ray binaries known to exhibit X-ray absorption lines from highly
ionized (H- and He-like) Fe and other metals, ordered by increasing
orbital period. The source name is given in column 1. The orbital
period taken from \citet[][ see references therein]{ritter03aa}
catalog, unless otherwhise mentioned, is given in column 2. Column 3
indicates if dips (D) and eclipses (E) are observed from the
source. The system inclination, $i$, is given in column 4. It is
deduced from the presence of dips or eclipses after
\citet{frank87aa}, unless otherwhise mentioned. The H-like and/or
He-like ions detected in each source are listed in column 5, and the
corresponding observatory and references used are given in column 6
and 7.
\nocite{1254:boirin03aa} 
\nocite{1658:sidoli01aa} 
\nocite{1624:parmar02aa}
\nocite{1655:orosz97apj}
\nocite{kuulkers98apj} 
\nocite{1655:orosz97apj}
\nocite{1655:ueda98apj} 
\nocite{1655:yamaoka01pasj}
\nocite{cirx1:kaluzienski76apjl}
\nocite{cirx1:shirey99apjb} 
\nocite{cirx1:brandt96mnras} 
\nocite{cirx1:brandt00apjl} 
\nocite{gx13:schnerr03aa}
\nocite{gx13:ueda01apjl} 
\nocite{gx13:sidoli02aa}
\nocite{gx13:corbet03apj}
\nocite{1915:mirabel94nature} 
\nocite{1915:kotani00apj} 
\nocite{1915:lee02apj}

}

\begin{tabular}{lllllll}

\hline
\hline
\noalign {\smallskip}
Source & P$_{\rm orb}$ (h)& Dips & $i$ (\degree) & H- or He-like ions & Observatory & References \\
      
\noalign {\smallskip}
\hline

\nineteen & 0.8 &  D    & 60--80 & Fe  & XMM & {\tiny This work}\\
\twelve & 3.9 &   D    & 60--80 & Fe  & XMM& {\tiny Boirin \& Parmar 2003}\\
\mxb & 7.1 &   D, E & $\sim$80 & O, Ne, Fe & XMM & {\tiny Sidoli et al. 2003}\\
\bigdip & 20.9 &  D    & 60--80 & Fe  & XMM & {\tiny Parmar et al. 2002}\\
\gro~\numaup & 62.9~\numaup & D~\numaup & $69.5 \pm 0.3$~$^2$ & Fe & ASCA & {\tiny Ueda et al. 1998, Yamaoka et al. 2001} \\
\cir & 398~\numaup & D~\numaup & high?~\numaup & Ne, Mg, Si, S, Fe  & {\it Chandra} & {\tiny Brandt \& Schulz 2000}~\numaup\\
\gx &   602~\numaup  & no  & low?~\numaup & Ca, Fe, Ni & ASCA, XMM & {\tiny Ueda et al. 2002, Sidoli et al. 2002}\\
\grs~$^1$ & 804 & no  & $\sim$70~\numaup & Ca, Fe, Ni & ASCA, {\it Chandra} & {\tiny Kotani et al. 2000, Lee et al. 2002} \\
\hline
\noalign {\smallskip}
\multicolumn{7}{l}{\numbup Microquasar}\\
\multicolumn{7}{l}{\numbup Orosz \& Bailyn 1997}\\
\multicolumn{7}{l}{\numbup Kuulkers et al. 1998}\\
\multicolumn{7}{l}{\numbup Kaluziensly et al. 1976}\\
\multicolumn{7}{l}{\numbup complex and irregular dipping behaviour (see, e.g., Shirey et al. 1999)}\\
\multicolumn{7}{l}{\numbup high $i$ suggested by spectral properties (Brandt et al. 1996)}\\
\multicolumn{7}{l}{\numbup braod P Cygni X-ray lines}\\
\multicolumn{7}{l}{\numbup Corbet 2003}\\
\multicolumn{7}{l}{\numbup low $i$ suggested by the high radio to X-ray flux ratio, in the framework of a Doppler boosting model (Schnerr et al. 2003)}\\
\multicolumn{7}{l}{\numbup $i$ inferred assuming the superluminal jets are
perpendicular to the disk (Mirabel \& Rodriguez 1994)}\\

\label{tab:sources}
\end{tabular}
\end{center}
\end{table*}

Table~\ref{tab:sources} lists the X-ray binaries where X-ray
absorption lines from highly ionized Fe or other metals have been
reported. We note that
\cir\ differs from the other sources by showing broad X-ray lines with
P Cygni profiles. The eclipsing, dipping and bursting LMXB \exo, has
shown X-ray absorption lines from H- and He-like ions, but these lines
differ from those found in the other sources in being redshifted,
detected during X-ray bursts and attributed to the neutron star
photosphere \citep{0748:cottam02nat}. Therefore, we did not include
\exo\ in the Table. Note however that the presence of an highly
ionized plasma is nevertheless revealed by the detection of emission
lines and absorption edges during the non-bursting emission of \exo\
\citep{0748:cottam01aa,0748:jimenez03apj}, as is also true in the
accretion disk corona source \eighteentt\ \citep{1822:cottam01apjl}.
The most obvious property probably shared by the sources listed in
Table~\ref{tab:sources} is to be viewed at high inclination. This
condition is strengthened by the discovery of absorption lines in the
dipping source \nineteen\ reported here. Among the eight sources, only
one, \gx, was suggested to have a low inclination, because it shows an
unusually high radio to X-ray flux ratio, that can be explained by
involving Doppler boosting of the radio emission, which implies that
the radio jets point towards the observer and thus, that the source is
seen close to pole-on \citep{gx13:schnerr03aa}. All the other sources
show strong evidence for being close to edge-on. This shows that
inclination angle is important in determining the strength of these
absorption features, which implies that the absorbing material is
distributed in a cylindrical, rather than a spherical geometry, around
the compact object. The azimuthal symmetry is implied by the lack of
any orbital dependence of these features \citep[e.g.,
][]{1658:sidoli01aa,1254:boirin03aa}, apart during dipping, where the
absorbing material seems to be less ionized (see below).  Thus, the
highly ionized material responsible for the absorption lines is likely
to be related to the accretion disk. Such a highly ionized plasma may
be common to systems accreting via a disk, but its observation made
easy when the system is highly inclined.

\citet{1254:boirin03aa} noticed that the ratio of \fetsix\ (H-like) to
\fetfive\ (He-like) line \ews\ was higher in \twelve\ than in \mxb\
and \gx, suggesting that the material responsible for the lines was
more strongly ionized in \twelve. The authors proposed that the
difference could be related to the overall size of the Roche lobe
around the compact object into which the accretion disk must
fit. Systems with shorter orbital periods are expected to have smaller
accretion disks, and the obscuring material may be expected to be more
photo-ionized in smaller systems with the same luminosities. 
As \nineteen\ has the shortest
orbital period and similar luminosities to the other systems
(see Table~\ref{tab:sources}), but has an absorbing
material less strongly ionized than in \twelve, this hypothesis seems
to be excluded.

Narrow absorption lines are detected in the EPIC pn persistent
emission of \nineteen\ at 6.65 and 6.95~keV and identified with \ka\
resonant absorption lines from \fetfive\ and \fetsix,
respectively. Since both \fetfive\ and \fetsix\ absorption features
are detected, some physical parameters of the plasma responsible for
the lines can be estimated. The column density of each ion can first
be estimated from the \ew\ of the corresponding absorption line, using
the relation quoted e.g., by \citet[][ see also references
therein]{1915:lee02apj} linking the two quantities, which is valid if
the line is unsaturated and on the linear part of the curve of growth,
which is verified in the case of \nineteen. The ratio between
\fetfive\ and \fetsix\ column densities can then be used to estimate
the photo-ionization parameter, $\xi$, using the calculations of
\citet{kallman01apjs}. Following this approach, we derive column
densities of $3.4 \times 10^{17}$~cm$^{-2}$ and $6.6 \times
10^{17}$~cm$^{-2}$ for \fetfive\ and \fetsix, respectively, and
estimate $\xi$ to be $10^{3.92}$ \xiunit\ during the persistent
emission.  An absorption line consistent with \fetfive\ \ka\ is also
detected in each intensity-selected dipping spectrum, and upper-limits
have been set on the \ew\ of a \fetsix\ \ka\ for each of these
spectra. Thus, we also estimate $\xi$ to be $\approxlt10^{3.83}$,
$\approxlt10^{3.68}$ and $\approxlt10^{3.68}$ \xiunit, during shallow,
intermediate and deep dipping, respectively. These values are
consistent with a decrease in ionization as dipping activity
increases.  This is again consistent with the presence of cooler
material in the line of sight during dipping.  We note that
\citet{1624:parmar02aa} also found evidence from the ratio of
\fetfive/\fetsix\ absorption line depths for the presence of
additional cooler material in the line of sight in another dip source,
\bigdip, during dipping.

\section{Conclusion}

We have reported the discovery of narrow absorption lines at 6.65 and
6.95~keV, consistent with resonant \fetfive\ and
\fetsix\ \ka\ transitions, in the persistent emission of the 
LMXB \nineteen. There is also marginal evidence for absorption
features consistent with \ion{Mg}{xii}, \ssixteen, \nitseven\ \ka\ and
\fetsix\ \kb\ transitions. Such absorption lines from highly ionized
ions are now observed in a number of LMXBs seen close to edge-on
($i\sim70$\degree), such as \nineteen.  This suggests that the highly
ionized plasma responsible for the absorption lines is related to the
accretion disk.  We have reported the dectection of the
\fetfive\ line in the dipping emission  of \nineteen, and the
upper-limits to the \fetsix\ column densities are consistent with a
decrease in the amount of ionization during dipping intervals.  This
implies the presence of cooler material in the line of sight during
dipping.  We have also reported the discovery of a 0.98~keV absorption
edge in the persistent emission spectrum.  The edge energy decreases
to 0.87~keV during deep dipping intervals. The detected feature may
result from edges of moderately ionized Ne and/or Fe with the average
ionization level decreasing from persistent emission to deep
dipping. This is again consistent with the presence of cooler material
in the line of sight during dipping.

\begin{acknowledgements}
Based on observations obtained with XMM-Newton, an ESA science mission
with instruments and contributions directly funded by ESA member
states and the USA (NASA).  L. Boirin acknowledges an ESA Fellowship.

\end{acknowledgements}


\bibliographystyle{aa}


\end{document}